\input{aipcheck}

\documentclass[
    ,final            
  ]
  {aipproc}

\layoutstyle{8x11double}

\begin{document}

\title{Cosmic Infrared Background from Early Epochs - Searching for Signatures of the First Stars}


\author{A. Kashlinsky}{
  address={SSAI and Code 665, Goddard Space Flight Center, Greenbelt, MD 20771, U.S.A.}
}

\begin{abstract}
Cosmic infrared background (CIB) contains emission from epochs
inaccessible to current telescopic studies, such as the era of the
first stars.  We discuss theoretical expectations for the CIB
contributions from the early population of massive stars. We then
present the latest results from the ongoing project by our team
\cite{kamm1,kamm2,kamm3,kamm4} to measure/constrain CIB
fluctuations from early epochs using deep Spitzer data. The
results show the existence of significant CIB fluctuations at the
IRAC wavelengths (3.6 to 8 $\mu$m) which remain after removing
galaxies down to very faint levels. These fluctuations must arise
from populations that have a significant clustering component, but
only low levels of the shot noise. Furthermore, there are no
correlations between the source-subtracted IRAC maps and the
corresponding fields observed with the {\it HST} ACS at optical
wavelengths. Taken together, these data imply that 1) the sources
producing the CIB fluctuations are individually faint with
$S_\nu<$ a few nJy at 3.6 and 4.5 $\mu$m; 2) are located within
the first 0.7Gyr (unless these fluctuations can somehow be
produced by - so far unobserved - local galaxies of extremely low
luminosity and with the unusual for local populations clustering
pattern), 3) they produce contribution to the net CIB flux of at
least 1-2 nW/m$^2$/sr at 3.6 and 4.5 $\mu$m and must have
mass-to-light ratio significantly below the present-day
populations, and 4) they have angular density of $\sim$ a few per
arcsec$^2$ and are in the confusion of the present day
instruments, but can be individually observable with {\it JWST}.
\end{abstract}

\maketitle


\subsection{Introduction}

Cosmic infrared background (CIB) contains emissions produced
during the entire history of the Universe, including those arising
from the epochs of the first stars (see \cite{review} for review).
Isolating the part of the CIB coming mostly from the first stars
epochs would provide important information about the emissions,
evolution and contents of this era. This paper reviews the recent
measurements done by our team in deep Spitzer IRAC data
(Kashlinsky, Arendt, Mather \& Moseley - herafter KAMM,
\cite{kamm1,kamm2,kamm3,kamm4}) and their interpretation.

The paper is structured as follows: it starts with a brief
discussion of the theoretical motivation for our measurements,
then presents a summary of the most recent results at the IRAC
Channels 1 (3.6 $\mu$m) and 2 (4.5 $\mu$m). Our most recent
analysis \cite{kamm4} showed that in the fields observed at {\it
both} IRAC and optical ACS {\it HST} bands there are no
correlations between the source-subtracted IRAC and the ACS source
catalog maps. The absence of these correlations would place the
sources producing these fluctuations within the first 0.7 Gyr
unless they originate in the more local, but extremely low
luminosity galaxies that somehow escaped ACS detection. Finally,
we discuss the data requirements on the nature of the populations
producing these fluctuations: they have to originate from very
faint sources, located within the first Gyr of the Universe's
evolution. These require them to have mass-to-light ratio
significantly below the present-day stellar populations and their
number density is such that they are well within the confusion of
the present instruments.

\subsection{Theoretical expectations}

An early generation of massive stars may have produced a
non-negligible contribution to the CIB at the near-IR wavelengths.
The net amount of that contribution can account for the entire
claimed CIB excess (see review \cite{review} for summary of
measurements up to 2005) over the contribution due to the observed
galaxy counts between 1 and 4 $\mu$m if only $\sim2-4\%$ of the
baryons went through these early stars \cite{grb}. The energy
spectrum of the resultant CIB would be cut off below the Lyman
limit at the redshift of these stars ($\sim 1\mu$m) and the
fraction of the baryons locked in them would decrease linearly
with the observed CIB excess flux. The latter would have to be
below $\sim 0.2-0.5\%$ if the NIR CIB excess is as low as the
NICMOS measurements \cite{thompson} suggest.

\begin{figure}
  \includegraphics[height=.225\textheight]{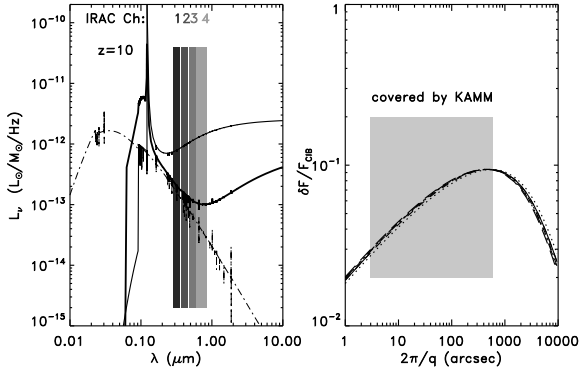}
\caption{{\bf Left}: SED from Pop~III system is shown for $z$=10.
The two solid lines are drawn from \cite{santos} for the case when
re-processing of the radiation takes place in the gas inside the
nebula and outside in the IGM. Thin dashed-dotted line shows the
intrinsic SED prior to the emission re-processing by gas. {\bf
Right}: Lines show the fluctuations expected for concordance
$\Lambda$CDM model at $z$=5 to 20 for the parameters described in
the text. Shaded area shows the angular scales probed by KAMM.
\label{fig:theory}}
\end{figure}

The left panel of Fig. \ref{fig:theory} shows a template of early
star emission at $z=10$ from \cite{santos}. The IRAC filters are
marked with shaded areas; the figure shows that, although the IRAC
bands are not centered at the peak of the emission produced by
Ly-$\alpha$, their wavelengths cover the region containing much of
the cumulative emission. This emission is likely to be reprocessed
by the surrounding gas resulting in the model-dependent
uncertainty at least as large as shown in the figure. Note that
the color of the emission at IRAC channels, while approximately
flat, can be either slowly or slowly decreasing with wavelength.

The first stars should also have left a distinct imprint in the
CIB anisotropies, measuring which could give important information
about that era and its constituents \cite{cooray,kagmm}. There are
intuitive reasons why the anisotropies can be significant: 1)
first stars, if massive, emitted a factor $\sim 10^5$ more light
per unit mass than the present-day stellar populations, 2) their
relative fluctuations would be larger because they span a
relatively narrow time-span in the evolution of the Universe, and
3) they formed (presumably) at the high peaks of the underlying
density field, which would amplify their clustering properties.

The CIB fluctuations, $\delta F$, on scale $2\pi/q$ are related to
the 2-D power spectrum, $P_2(q)$, of the diffuse light via $\delta
F \simeq [q^2P_2(q)/2\pi]^{1/2}$. They are related to the CIB
produced  by them, $F$, and the underlying 3-D power spectrum of
the sources' clustering, $P_3(k)$, via the Limber equation which
approximates as $\delta F \simeq F \Delta(qd_A^{-1}(z))$, where
$\Delta^2 \equiv \frac{1}{2\pi^2}\frac{k^2P_3(k)}{\Delta t}$ is
the mean square fluctuation in the source counts over a cylinder
of radius $\sim k^{-1}$ and length $c\Delta t$ assuming the
sources span cosmic time $\Delta t$. The right panel of Fig. 1
shows the typical relative fluctuation in the CIB produced by the
early sources over $z$ from 6 to 20 assuming a (realistic)
``toy"-model with the concordance $\Lambda$CDM initial conditions
where the amplification due to biasing is $A\simeq 0.3 (1+z)
\sqrt{\Delta t({\rm Gyr})}$. The figure shows the distinct peak at
$\simeq 0.2-0.3$ degrees which, for CDM class of models,
corresponds to the horizon at the matter-radiation equality
redshifted to $z$. The shaded area shows the range of angular
scales covered so far by the KAMM measurements. At these scales
one can expect fluctuations of no more than $3-10\%$, so the CIB
fluctuation level of $\delta F\sim 0.1$ nW/m$^2$/sr would require
CIB of at least $F\sim (1-2)$ nW/m$^2$/sr.

\subsection{CIB fluctuations from Spitzer deep images}

The KAMM methodology is summarized in \cite{kamm1,kamm2}. Briefly,
the steps are: 1) The maps are assembled using the least
calibration method of \cite{fixsen}. It solves for detector
properties at the time of observations and for frame-to-frame
variations, and does not remove large-scale structure except due
to (arbitrary) linear gradients. 2) The assembled IRAC maps are
first clipped of resolved sources via an iterative procedure which
removes all pixels (with a certain mask) with flux above a certain
threshold. 3) The clipped maps are then further cleaned of the
remaining emissions from resolved sources using an iterative
variant of the clean procedure \cite{clean}. 4) The clipping is
stopped so that enough pixels are left (in practice $> 70\%$) for
robust computation of the fluctuations spectrum via Fourier
transform. 5) When clipping is done to deeper levels, and
progressively smaller fractions of pixels remain, we compute
instead the correlation function, which is immune to masking
\cite{kamm1}. 6) The uncertainties on the resultant power spectra
(e.g. due to mask which destroys the strict orthogonality of the
basis functions) are verified via simulations. 7) It is verified
that, at the iterations used for measuring the final fluctuations,
the subtracted sources components do not correlate with the
residual diffuse emissions of the maps.

\begin{figure}
  \includegraphics[height=.175\textheight]{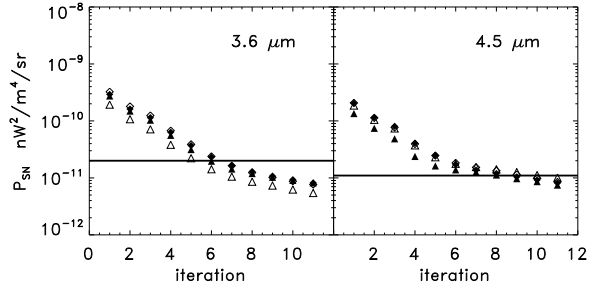}
  \caption{Shows decrease of the shot noise
contribution to the power spectrum vs the iteration number of the
KAMM source subtraction model. Horizontal lines show the levels
reached in \cite{kamm2}, which are a factor of $\sim 2$ below
those in \cite{kamm1}.  Triangles correspond to the HDFN region,
diamonds to CDFS region; open symbols to Epoch 1, filled to Epoch
2. \label{fig:sn}}
\end{figure}

\begin{figure}
  \includegraphics[height=.425\textheight]{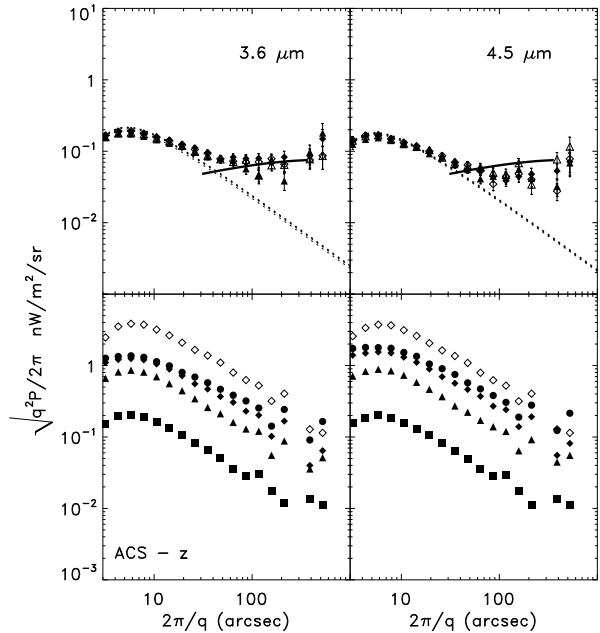}
  \caption{{\bf Top panels}: CIB
fluctuations from the source-subtracted maps from \cite{kamm4} at
the shot-noise levels of \cite{kamm2}. Same symbol notation as in
Fig. \ref{fig:sn}. Dotted lines show the shot-noise contribution.
Solid line shows the slope of sources at high-$z$ with the
$\Lambda$CDM model spectrum of the same amplitude at 3.6 and 4.5
$\mu$m. {\bf Lower panels}: CIB fluctuations due to ACS galaxies
for the $972\times 972$ 0.6$^{\prime\prime}$ pixel field at
HDFN-Epoch2 region for the ACS z-band. Right and left panels show
the spectrum for the ACS source maps convolved with the IRAC 3.6
 and 4.5 $\mu$m beams. Filled circles correspond to ACS
galaxies fainter than $m_0=21$ with the mask defined by the
clipping. Filled diamonds, triangles and squares correspond to
fluctuations produced by sources fainter than $m_0+2,m_0+ 4,
m_0+6$. Open diamonds show the fluctuations produced by galaxies
fainter than $m_{\rm AB}$=23 when the clipping mask is not
applied; the symbols show that such galaxies were effectively
removed from the ACS maps by clipping alone done in KAMM.
\label{fig:kamm}}
\end{figure}

As of now we have analyzed results from five deep exposure regions
in various parts of the sky \cite{kamm1,kamm2}. The latest
results, shown here in Figs. \ref{fig:sn},\ref{fig:kamm}, are from
analysis of four independent regions mapped in the course of the
GOODS project observations. They cover $\simeq 10^\prime$ on the
side and have total exposure time of $\sim 23-25$ hours per pixel.
The power spectrum of the diffuse emission in the source
subtracted maps is made up of two components: 1) small scales are
dominated by the shot-noise from the remaining sources, and 2)
larger scales are dominated by the component of CIB fluctuations
arising from clustering of the contributing sources. As we clean
maps of progressively fainter sources, the shot-noise level
decreases as shown in Fig. \ref{fig:sn}. At the same level of the
shot noise all five regions exhibit the same large-scale
fluctuations consistent with their cosmological origin. The figure
also shows the shot-noise levels we stop at when presenting the
results; they are limited by the floor imposed by the instrument
noise levels. Comparing the shot-noise levels with those computed
from the observed galaxy counts tells us that individual galaxies
have been eliminated to AB magnitudes $>26$.

The upper panels in Fig. \ref{fig:kamm} show the CIB fluctuations
from the GOODS data at 3.6 and 4.5 $\mu$m. Dotted lines show the
shot-noise. We demonstrated that these fluctuations do not arise
from the various Solar System and Galactic foregrounds and are not
produced by the instrument systematics \cite{kamm1,kamm2}. The
fluctuations also correlate well with maps at the longest IRAC
channels (5.8 and 8 $\mu$m), although at these channels we do not
recover CIB fluctuations with the same S/N due to the larger
instrument noise and possible cirrus pollution at 8 $\mu$m. The
clustering component dominates fluctuations at $>20-30 ^\prime$.
The solid line shows the fluctuation spectrum due to the
concordance $\Lambda$CDM model at high $z$ assumed for simplicity
to have the same amplitude at 3.6 and 4.5 $\mu$m. The figure shows
that the signal is well described by the $\Lambda$CDM model with
sources at high-$z$ and the color of these sources is
approximately flat at the two IRAC bands.

\subsection{Cross-correlating NIR and optical data}

The GOODS fields have also been observed with the {\it HST} ACS
instrument at optical wavelengths from $\sim 0.4$ to 0.9 $\mu$m
reaching source detection levels fainter than 28 AB mag. If these
fluctuations arise from local populations there should be a strong
correlation between the source-subtracted IRAC maps and the ACS
sources. Conversely, there should be no such correlations if the
KAMM signal arises at at epochs where the Lyman break (at rest
$\sim 0.1\mu$m) gets shifted passed the longest ACS z-band at
$\simeq 0.9\mu$m. To test for this we \cite{kamm4} have
constructed synthetic maps, overlapping with the GOODS fields,
using sources in the ACS B,V,i,z bands from the ACS sources
catalog \cite{goods}. These maps were then convolved with the IRAC
3.6 and 4.5 $\mu$m beams. Finally, we applied the clipping mask
from the IRAC maps and computed the fluctuations spectrum produced
by the ACS sources and their correlations with the IRAC-based
maps.

\begin{figure}
  \includegraphics[height=.2075\textheight]{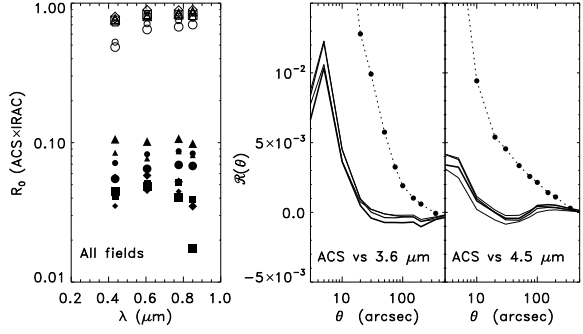}
\caption{{\bf Left}: Correlation coefficient between
clipped/masked ACS and KAMM data. Large and small symbols
correspond to the IRAC Ch 1 and Ch 2. Open symbols correspond to
correlations with the maps of the removed sources and filled
symbols with the residual KAMM maps. Circles, diamonds, triangles,
squares correspond to HDFN-E1, HDFN-E2, CDFS-E2, CDFS-E2. {\bf
Right}:  The various thickness solid lines show the dimensionless
correlation function between the diffuse light in the ACS and KAMM
maps for $B,V,i,z$-bands. Dotted line shows the dimensionless
correlation function of the KAMM maps, $C_{\rm
KAMM}(\theta)/\sigma_{\rm KAMM}^2$, which remains positive out to
$\sim 100^{\prime\prime}$ and is better viewed when presented in
log-log plots as in Fig. SI-4 of \cite{kamm1}.  \label{fig:xcor}}
\end{figure}

The lower panels in Fig. \ref{fig:kamm} show the spectrum of
fluctuations produced by the ACS z-band sources. It is clear from
the figure that the ACS sources produce a completely different
spectrum of the fluctuations than found by KAMM. This alone shows
that they cannot produce the KAMM fluctuations signal shown in the
upper panels. In fact, their power spectrum is very similar to
that of the CIB fluctuations found in the deep 2MASS data, which
arise from galaxies at $z>1$ \cite{2mass}. Fig. 2e of ref.
\cite{2mass} shows the gradual evolution of the power spectrum
spectral slope towards deeper cut 2MASS data which is consistent
with the evolution of the various slices of the ACS sources in the
lower panels of Fig. \ref{fig:kamm}.

In ref. \cite{kamm4} we compute both the correlation coefficients,
$R_0\!=\!\langle \delta F_{\rm ACS}\delta F_{\rm
KAMM}\rangle/\sigma_{\rm ACS}\sigma_{\rm KAMM}$, and the full
cross-correlation matrix, $C(\theta)\!=\!\langle \delta F_{\rm
ACS}(\vec{x})\delta F_{\rm KAMM}(\vec{x}+\vec{\theta})\rangle$
between the ACS and KAMM maps. In this representation, the
contributions of any white noise (such as shot noise and/or
instrument noise) component to $C(\theta)$ drop off very rapidly
outside the beam and for the IRAC 3.6 $\mu$m channel contribute
negligibly to the correlation function at $\theta$ greater than a
few arcsec. This is best seen from the analog of the correlation
coefficient at non-zero lag, defined as
$R(\theta)\!\equiv\!C(\theta)/\sigma_{\rm ACS}\sigma_{\rm KAMM}$.
The mean square fluctuation on scale $\theta$ is given by the
integral $\langle \delta F(\theta)^2\rangle$=$\frac{2}{\theta^2}
\int_0^\theta C(\theta^\prime)\theta^\prime d\theta^\prime$.
Hence, we evaluated from $R(\theta)$ a related quantity ${\cal R}
(\theta)$=$\frac{2}{\theta^2} \int_0^\theta
R(\theta^\prime)\theta^\prime d\theta^\prime$, shown in the right
panels in Fig. \ref{fig:xcor}.

There are excellent correlations (shown with open symbols) between
the ACS source maps and the sources {\it removed} by KAMM prior to
computing the remaining CIB fluctuations in Fig. \ref{fig:kamm}.
However, practically no correlations remain with the
source-subtracted IRAC maps and the correlations are negligible
outside the IRAC beam. This means that, at most, the remaining ACS
sources contribute to the shot-noise levels in the residual KAMM
maps (as discussed by \cite{kamm3}), but not to the large scale
correlations in Fig. \ref{fig:kamm}.

\subsection{The nature of the sources.}

Any interpretation of the KAMM results must reproduce the
following major aspects: $\bullet$ The sources in the KAMM data
were removed to a certain (faint) flux limit, so the CIB
fluctuations arise in populations with magnitudes fainter than the
corresponding magnitude limit. $\bullet$ These sources must
reproduce the excess CIB fluctuations by KAMM on scales $>
0.5^\prime$, {\it both} their amplitude and their spatial
spectrum. $\bullet$ The populations fainter than the above
magnitude limit must account not only for the correlated part of
the CIB, but - equally importantly - they must reproduce the (low)
shot-noise component of the KAMM signal, which dominates the power
at $<$0.5$^\prime$. $\bullet$ The data show the absence of
correlations between the source-subtracted IRAC maps and the same
area mapped by ACS. $\bullet$ The CIB fluctuations signal is
measured from the source-subtracted IRAC maps at all four channels
(although not with the same S/N at the longest wavelengths) and
has color which is approximately flat with wavelength across the
IRAC bands.

We briefly discuss the constraints in the above order:

1) {\bf Magnitude limits}. The nominal limit above which sources
have been removed in the KAMM analysis is $m_{\rm AB} \simeq 26$
at 3.6 and 4.5 $\mu$m and this by itself implies that the detected
CIB fluctuations arise from fainter systems. At this magnitude one
is already at the confusion limit for the IRAC beam at 3.6 $\mu$m,
so fainter galaxies can be excised only by removing significantly
more pixels. This magnitude limit corresponds to $6 \times 10^8
h^{-2} L_\odot$ emitted at 6000 \AA\ at $z$=5. A significant
fraction of galaxies were thus removed from the data by KAMM even
at $z\geq 5$ and the detected CIB fluctuations must be explained
by still fainter and more distant systems.

 2) {\bf Clustering component} The relative CIB fluctuations
produced by these sources are at most 10\% (see Fig.
\ref{fig:theory} and \cite{kamm3}). Fig. \ref{fig:kamm} shows that
the clustering strength at $\geq 1^\prime$ requires $\delta F_{\rm
CIB}\sim 0.1$ nW/m$^2$/sr. This meaning that the net CIB from
sources contributing to the KAMM signal at 3.6 $\mu$m is $>$1-2
nW/m$^2$/sr.

3) {\bf Shot noise constraints}. The amplitude of the shot-noise
power gives a particularly strong indication of the epochs of the
sources contributing to the KAMM signal \cite{kamm3}. This can be
seen from the expressions for the shot noise $P_{\rm SN} =
f(\bar{m}) F_{\rm tot}(>\!m_{\rm lim})$, where $f(m)$ is the flux
of the source of magnitude $m$ and $F_{\rm tot}(>m)$ is the net
CIB flux produced by sources fainter than $m$ \cite{review}. Fig.
\ref{fig:kamm} shows the shot noise amplitude evaluated at 3.6 and
4.5 $\mu$m is $P_{\rm SN} \simeq (2,1) \times 10^{-11}$
nW$^2$/m$^4$/sr. Above it was shown that the sources contributing
to the fluctuations must have CIB flux greater than a few
nW/m$^2$/sr and combining this with the values for $P_{\rm SN}$
shown leads to these sources having typical magnitudes $m_{\rm AB}
> 29-30$ or individual fluxes below a few nJy.

4) {\bf No correlations with optical ACS data}. Our results show
that the source-subtracted CIB fluctuations in deep Spitzer images
cannot originate in the optical galaxies detected in the GOODS ACS
data. While the $B, V, i, z$-band galaxies are well correlated
with those seen {\it and removed} in the 3.6 and 4.5 $\mu$m data,
they correlate poorly with the the residual 3.6 and 4.5 $\mu$m
background emission. These galaxies also exhibit a very different
spatial power spectrum than the KAMM maps and the amplitude of
their fluctuations is generally also low. Whatever sources are
responsible for the KAMM fluctuations, they are not present in the
ACS catalog. Since the ACS galaxies do not contribute to the
source-subtracted CIB fluctuations, the latter must arise at $z>7$
as is required by the Lyman break at rest $\sim 0.1 \mu$m getting
redshifted past the ACS $z$-band of peak wavelength $\simeq 0.9
\mu$m. This would place the sources producing the KAMM signal
within the first 0.7 Gyr. If the KAMM signal were to originate in
lower $z$ galaxies which escaped the ACS GOODS source catalog
because they are below the catalog flux threshold, they would have
to be extremely low-luminosity systems ($< 2\times 10^7
h^{-2}L_\odot$ at $z$=1) and these galaxies would also have to
cluster very differently from their ACS counterparts as Fig.
\ref{fig:kamm} shows.

5) {\bf Color}. Color gives additional information on the nature
of the sources producing these fluctuations. Fig. \ref{fig:kamm}
shows that the clustering of the populations producing the KAMM
signal at 3.6 and 4.5 $\mu$m is well described by the concordance
$\Lambda$CDM model with sources at high $z$ with the color
approximately expected from populations described by SED of the
type in Fig. \ref{fig:theory}. The signal is not as clear at the
two longest IRAC bands, but the data at all four channels
correlate meaning that the same sources contribute to all the
wavelengths \cite{kamm1}. The color at shorter wavelengths can
prove an interesting diagnostic \cite{thompson2}, but requires
additional assumptions on the epochs of the emissions (which
determine the location of the Lyman break) and that the
differently processed maps leave the same populations.

Additional information on the nature of the populations
responsible for these CIB fluctuations, can be obtained from the
fact that the significant amount of flux ($>$1-2 nW/m$^2$/sr)
required to explain the amplitude of the fluctuations must be
produced within the short time available at these high $z$ (cosmic
times $<$0.5-1 Gyr). This can be translated into the comoving
luminosity density associated with these populations, which in
turn can be translated into the fraction of baryons locked in
these objects with the additional assumption of their
$\Gamma\equiv M/L$ \cite{kamm3}. The smaller the value of
$\Gamma$, the fewer baryons are required to explain the CIB
fluctuations detected in the KAMM studies. It turns out that in
order not to exceed the baryon fraction observed in stars, the
populations producing these CIB fluctuations had to have $\Gamma$
much less than the solar value, typical of the present-day
populations \cite{kamm3}. This is consistent with the general
expectations of the first stars being very massive.

For a typical Population III system with an SED such as shown in
Fig. \ref{fig:theory}, one expects $\Gamma \sim
10^{-3}-10^{-2}\Gamma_\odot$ at the IRAC bands. Because the (low)
shot-noise values at 3.6 and 4.5 $\mu$m imply that the individual
sources have flux below a few to 10 nJy, the bulk of these
populations had to have only at most a few times $10^5M_\odot$ of
stellar material if they are to explain the KAMM fluctuations
\cite{kamm3}. Such sources would be below the detection threshold
of the current high-$z$ Lyman-break searches.

Finally, in order to detect the faint sources responsible for the
CIB fluctuations with fluxes below a few nJy embedded in the
underlying sea of emissions, their individual flux must exceed the
confusion limit \cite{kamm3}. If such sources were to contribute
to the CIB required by KAMM data, at 3.6 and 4.5 $\mu$m they had
to have the average surface density of $\bar{n} \sim F_{\rm
CIB}^2/P_{\rm SN} \sim 5 \; {\rm arcsec}^{-2}$. In order to avoid
the confusion limit and resolve these sources individually at,
say, 5-sigma level ($\alpha=5$) one would need a beam of the area
$\omega_{\rm beam} \leq \alpha^{-2}/\bar{n} \sim 5\times
10^{-3}{\rm arcsec}^2$ or of circular radius below $\sim$0.04
arcsec. This is clearly not in the realm of the currently operated
instruments, but the {\it JWST} could be able to resolve these
objects given its sensitivity and resolution.

{\bf Acknowledgements} I thank my collaborators, Rick Arendt, John
Mather and Harvey Moseley for many contributions to the KAMM
results, and the NSF AST-0406587 grant for support.

\bibliographystyle{aipproc}   

\end{document}